\documentclass[10pt,leqno]{amsart}
\usepackage{graphicx}
\usepackage{indentfirst,csquotes}

\usepackage{amssymb,amsthm,amsmath}
\usepackage{xcolor,paralist,hyperref,titlesec,fancyhdr,etoolbox}

\titleformat{\section}[display]{\normalfont\huge\bfseries\centering}{\centering\chaptertitlename\thechapter}{10pt}{\Large}
\titlespacing*{\section}{0pt}{0ex}{0ex}

\hypersetup{ colorlinks=true, linkcolor=black, filecolor=black, urlcolor=black }

\usepackage{lipsum}

\begin{document}
\title{Mid-infrared optical properties of non-magnetic-metal/C\MakeLowercase{o}F\MakeLowercase{e}B/M\MakeLowercase{g}O heterostructures} 
\author[Flores-Camacho $et~al$]{J.M. Flores-Camacho$^1$, B. Rana$^2$, R.E. Balderas-Navarro$^1$, A. Lastras-Mart\'{\i}nez$^1$, Y. Otani$^{3,}$$^4$, and J. Puebla$^4$}

\address{$^1$Instituto de Investigaci\'on en Comunicaci\'on \'Optica. Universidad Aut\'onoma de San Luis Potos\'\i, \'Alvaro Obreg\'on 64, 78000 San Luis Potos\'\i, Mexico}
\address{$^2$ ISQI, Faculty of Physics, Adam Mickiewicz University, 
Uniwersytetu Poznanskiego 2, Poznan 61–614, Poland}
\address{$^3$ Institute for Solid State Physics, The University of Tokyo, 
Kashiwa, Chiba 277-8581, Japan}
\address{$^4$ CEMS, RIKEN, 2-1 Hirosawa, Wako, Saitama 351-0198, Japan}
\email{(jmfc): manuel.camacho@uaslp.mx; (rbn): raul.balderas@uaslp.mx; (jp): jorgeluis.pueblanunez@riken.jp}
\maketitle

\let\thefootnote\relax
\footnotetext{} 

\begin{abstract}
We report on the optical characterization of non-magnetic metal/ferromagnetic (Co$_{20}$Fe$_{60}$B$_{20}$)/MgO heterostructures and interfaces by using mid infrared spectroscopic ellipsometry at room temperature. We extracted for the mid-infrared range the dielectric function of Co$_{20}$Fe$_{60}$B$_{20}$, that is lacking in literature, from a multisample analysis. From the optical modelling of the heterostructures we detected and determined the dielectric tensor properties of a two-dimensional gas (2DEG) forming at the non-magnetic metal and the CoFeB interface. These properties comprise independent Drude parameters for the in-plane and out-of plane tensor components, with the latter having an epsilon-near-zero frequency within our working spectral range. A feature assigned to spin-orbit coupling (SOC) is identified. Furthermore, it is found that both, the interfacial properties, 2DEG Drude parameters and SOC strength, and the apparent dielectric function of the MgO layer depend on the type of the underlying nonmagnetic metal, namely, Pt, W, or Cu. The results reported here should be useful in tailoring novel phenomena in such types of heterostructures by assessing their optical response noninvasively, complementing existing characterization tools such as angle-resolved photoemission spectroscopy, and those related to electron/spin transport. 
\end{abstract} 

\bigskip

\section*{Introduction}
CoFeB is a common magnetic layer in spintronic devices. In particular, when sandwiched between MgO and non-magnetic heavy metals (Pt, W, Ta), the combination of bulk and interfacial spin–orbit coupling (SOC) together with the formation of an interfacial two-dimensional electron gas (2DEG) result in intriguing magnetization phenomena \cite{Torrejon14natc, Tacchi17prl, Zhang15apl}. In this regard, the nature of the 2DEG and the interfacial SOC at nonmagnetic metal (NM)/ferromagnetic metal (FM) interfaces is still a very active topic of research~\cite{nakayama18sciAdv}, that can greatly benefit from alternative experimental techniques to
transport measurements and magnetic imaging. The interplay between the formation of a 2DEG and SOC is not only of particular importance from the fundamental point of view but also plays a crucial role in device performance~\cite{Dolui, Zhu, Avci}. The formation of an interfacial 2DEG between metallic layers is
a consequence of strong interfacial hybridization, which in the case of heavy NM/FM interface, the understanding at the moment involves 3$d$-5$d$ hybridization, with important consequences for the technological relevant interfacial spin orbit torque~\cite{Zhu, Avci, Amin}. Hence, it is highly desirable to count on non-invasive
and quantitative experimental tools to detect the existence of both 2DEG and SOC, preferably in a single experimental technique. In this regard, optical probes are effective due to their versatility to operate in any ambient, both in-situ and ex-situ, in a non-invasive way. 

Here, we report the optical response of Co$_{20}$Fe$_{60}$B$_{20}$-based heterostructures at room temperature obtained by mid-infrared spectroscopic ellipsometry (IRSE), which is an optical technique that measures in a single shot relative changes of amplitude and phase, upon reflection, of a previously prepared polarized incident light. Its access to these two experimental parameters, amplitude and phase, permits the determination of complex quantities such as dielectric functions (DFs) and optical conductivities. IRSE is suitable to obtain information on vibrational modes and phonons of organic materials~\cite{Hinrichs} and crystals, respectively, and can be applied to characterize infrared plasmonic materials and interfacial properties of perovskites heterojunctions~\cite{Yazdi}. Infrared ellipsometry has been used to characterize ferromagnetic materials~\cite{Stewart}, elucidate origins of ferromagnetism~\cite{Fris}, and study band structure~\cite{VanDer}.  

Specifically, this paper addresses both, the optical response stemming solely from the CoFeB thin layer, and the understanding of IRSE spectra in terms of both epilayers and interfaces comprising the heterostructures, correlating them with signatures associated with interfacial 2DEG and SOC through a self-consistent analysis of their DFs.

The DF of the CoFeB material measured in the mid-infrared (MIR) range in this work, also covers a lacking part in the literature of the spectrum of CoFeB data, which should complement already existing relevant reports of spectroscopy ellipsometry studies in the near infrared and ultraviolet–visible (uv–vis) spectrum of samples based on CoFeB~\cite{Liang, Axel, Sharma, Sumi}, in which the DF was extracted from layers with different Co-Fe proportions and annealing temperature dependent states
of crystallinity. CoFeB nanocomposites for magnetic applications have also been studied by optical means~\cite{Kalashnikova, Kravets, Stas, Lys}, including an IR vibrational study of low-dimensional magnetic CoFeB-based materials~\cite{Domas}. In particular, Kravets $et~al$~\cite{Kravets}, extend their results for CoFe nanoparticles to the infrared where they detect an enlargement of magneto-optical Kerr rotation driven by the localized plasmon resonance of the composite. Enhancement of the Kerr rotation may be produced by an externally provided surface plasmon, as in the Au grating deposited on CoFeB~\cite{QWang}, by control of Kerr
null points~\cite{WZhang}, or by the epsilon-near-zero (ENZ) response of the material itself as reported for Weyl semimetals~\cite{JWu}. Spectroscopic ellipsometry (SE) in the infrared covers a spectral region where knowledge of optical conductivity, which is straightforwardly obtained from measurements of the complex DFs, is required to model magnetic field-dependent optical responses, but was, however, traditionally accessed by reflectivity alone~\cite{Kravets2}. 

The analysis of the optical properties of (Cu, Pt, W)/Co$_{20}$Fe$_{60}$B$_{20}$(1.5 nm)/MgO heterostructures brings information on the NM-dependent special properties, clearly different from those of the constitutive materials of the structure, of the interfacial 2DEG, which are proposed here as forming an anisotropic DF whose out-of-plane component contains an ENZ frequency. In our study, the ENZ has been used as a means for detection, due to a distinctive spectral feature, of the 2DEG, but we acknowledge that it may lead to more applications as in~\cite{JWu}. Additionally, we have detected that the MgO DF is affected by the type of NM metal in the structure in a manner that cannot be justified by crystal coherency alone. Thus, we believe our results will contribute to further understanding of the design of non-magnetic/magnetic heterostructures with specific SOC-driven phenomena such as direct and inverse spin Hall effect, Rashba–Edelstein effect and, ultimately, Dzyaloshinskii–Moriya interaction. We note that our study also aims at reigniting interest of studying the frequency dependent conductivity of magnetic samples by light, particularly in the infrared region, a research route that was perceived more than 50 years ago by Moriya and others~\cite{Moriya, Moriya2, Tsuchida}, with limited efforts thereafter~\cite{Hasegawa, Abraha}, particularly regarding experiments. 

The rest of the paper is organized as follows, section 2 describes the samples’ fabrication process, and the details of SE. The model for the IRSE analysis is discussed in section 3. Section 4 deals with the experimental results and discussion. Finally, the conclusions are outlined in section 5.

\section*{Experiment} 

Samples were grown by sputtering technique at room temperature onto thermally oxidized Si(100) substrate, consisting of NM(10)/Co$_{20}$Fe$_{60}$B$_{20}$(1.5)/MgO(2)/Al$_2$O$_3$(10) heterostructure layers, where NM, the non-magnetic metallic layer, stands for either W, Pt, or Cu in our study. The numbers in parentheses correspond to nominal layer thicknesses in nm. After the growth, the samples were annealed in the presence of a 0.6 T magnetic field along the normal axis to the film plane to induce interfacial perpendicular magnetic anisotropy. Additionally, a set of samples with a thicker CoFeB
layer and no MgO, i.e. SiO$_2$/Co$_{20}$Fe$_{60}$B$_{20}$(5)/Al$_2$O$_3$(10), SiO$_2$/Co$_{20}$Fe$_{60}$B$_{20}$(20)/Al$_2$O$_3$(10), and SiO$_2$/Ta(10)/Co$_{20}$Fe$_{60}$B$_{20}$(20)/Al$_2$O$_3$(10), were prepared as references to extract the CoFeB DF. 

For the optical characterization, SE measures the change in polarization state that a linearly polarized light incident at oblique angle experiences upon reflection off an optical medium of interest~\cite{Wee}. The incident light is linearly polarized to probe the three principal optical axes of the medium under study. The polarization state reflected by the medium is, in general, elliptical. For the analysis of the polarization states, the electric field of the light is mathematically decomposed into projections along and perpendicular to the plane of incidence of the light, $\mathbf{E}^i_p$ and $\mathbf{E}^i_s$ in figure 1, respectively; or simply $p$- and $s$- polarizations, respectively~\cite{Wee, Azzam, Fujiwara}. An ellipsometry experiment works by measuring the relative changes of amplitude and phase of the $p$- and $s$-polarizations~\cite{Azzam, Fujiwara} through the complex ratio $\rho$ of the oblique incidence reflection coefficients $r_p/r_s$ expressed as

\begin{equation}
    \rho = \frac{r_p}{r_s} = \tan\psi e^{i\Delta},
    \label{eq:rho}
\end{equation}
\noindent where $r_p$ ($r_s$) is the complex reflection coefficient of the whole stack (i.e., the Fresnel coefficients), for light polarized parallel (perpendicular) to the plane of incidence, $\psi$ is a measure of the relative intensities ratio and $\Delta$ is the relative wave phase shift~\cite{Azzam, Fujiwara}. We note its advantages over direct reflection and transmission experiments are double. First, as SE measures two parameters simultaneously, it provides access to the real and imaginary parts of the medium’s optical properties in a single experiment without mathematical post-treatment of the results. Second, SE has no necessity for reference measurements of intensities that can easily be altered by lamp instabilities or environmental changes, which are critical for infrared measurements. 

\begin{figure}
\centering
\includegraphics[width=7cm]{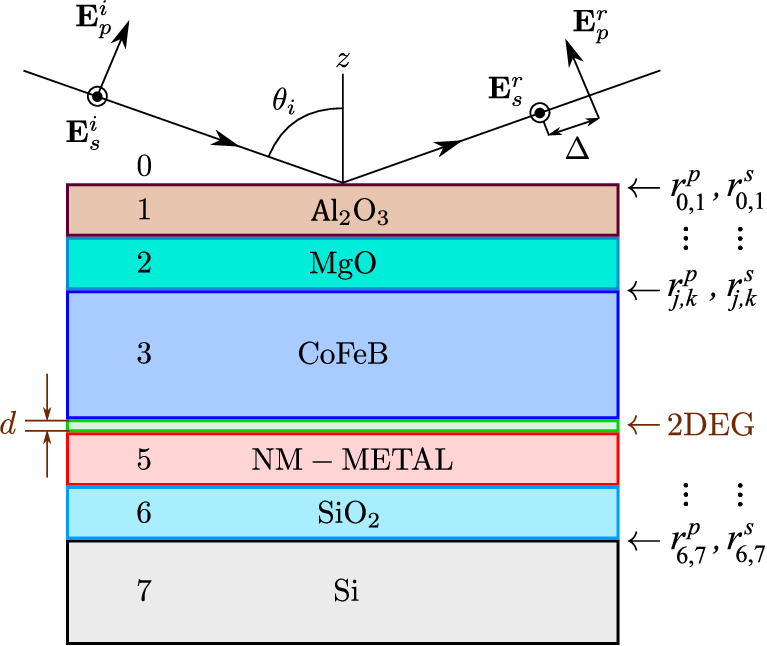}
\caption{Schematic diagram of the heterostructures grown on Si/SiO$_2$ substrates explored in this work. The non-magnetic (NM)-metal film can be either W, Cu or Pt. 
A two-dimensional electron gas (2DEG) with thickness $d$ is introduced in the stratified media accounting for the spin-orbit coupling. Ellipsometry measurements are performed with an angle of incidence $\theta_i$. $\mathbf{E}_s$ and $\mathbf{E}_p$ stand for perpendicular and parallel polarizations with respect to the plane of incidence, respectively. $\Delta$ is a measure for the relative retardance between $p$ and $s$ polarizations upon reflection coming from the whole structure's depth. 
The nominal thicknesses of the SiO$_2$, NM-Metals, CoFeB, MgO, and Al$_2$O$_3$ films are 700, 10, 1.5, 2 , and 10~nm, respectively.}    
\label{fig:SchematicSample}
\end{figure}

For this work, IRSE  was measured in the range of $\sim$50 to 800~meV ($\sim$25 to 1.55~$\mu$m) at different angles of incidence ($\theta_i$ in figure 1) using a commercial apparatus of the rotating compensator type (IR-VASE from J.A. Woollam Co.) in ambient atmosphere at room temperature. Before IRSE measurements, the samples' backsides were sandblasted to avoid partially or completely incoherent reflections. 

\section*{Model}

The experimental spectra are modeled by considering a stratified optical medium as the one in figure 1, for which interfacial reflection coefficients, film thicknesses, and complex DFs of each layer are taken into account to obtain total Fresnel coefficients for $s$ and $p$ polarizations separately~\cite{Azzam, Fujiwara}. The Fresnel coefficients are formed by considering the complex interfacial reflection amplitudes at each interface and a phase shift, $\delta$, induced by the optical path traversed by the wave at each layer (see supplementary material for the case of anisotropic/isotropic interface). The latter also serves for
correctly taking into account the attenuation at which the wave arrives at the next interface in the case of lossy media. Since the ellipsometry technique only has access to relative quantities, the separate outputs of the model are to be divided to obtain the ratio $\rho$ in eq.(1). Thus, $\psi=\tan^{-1}|r_p/r_s|$, and $\Delta=\arg(r_p)-\arg(r_s)$.

Simulations were performed with the aid of the software WVASE32 included in the commercial IRSE apparatus using an ad-hoc iterative procedure consisting on a combination of the built-in automated least mean square error procedure and manual parameterizations: it is reported that at some interfaces, for example, between different perovskites, new states appear that modify the optical response of the system as compared to the one expected by just mathematically stacking material layers~\cite{Asmara}. A systematic procedure to obtain the dielectric response of the layer containing such new state is demonstrated in~\cite{Asmara} for the isotropic case. In the present work, on the other hand, we propose that this new layer must be anisotropic. In such case the least-mean-squares automated routine is much more challenging in the setting of initial values than the isotropic one: it tends to fall on local minima instead of on the absolute minimum failing thus, to reproduce some experimental critical features, which we associate to the out-of-plane component of the mentioned new state dielectric response. Our approach to handle the initial settings, removing arbitrariness, relies on iterations in which variables’ correlations are quantified. For example, the matrix in table~S2 in the Supplementary Section shows high interdependence of of the parameters of the out-of-plane 2DEG layer DF, $\varepsilon_{zz}$  (highlighted in bold). We propose solving this issue by a qualitative uniqueness analysis as the one shown in 4.3. Measurement and analysis of several reference samples backed up this method.

As the number of parameters is large, the model must have the most considerable input data possible to leave only a few quantities as fitting parameters. In the present case, the DFs in the MIR range of Si~\cite{Herz}, SiO$_2$~\cite{Kitamura}, and Al$_2$O$_3$~\cite{Kisch} were taken from literature or the WVASE32 database, the CoFeB DF had to be separately determined from reference samples, and the NM DF had to be adapted using a percolation theory. Once these DF’s have been obtained they are used for the NM/CoFeB/MgO samples without further modifications. In this way, we are left with the properties of the 2DEG layer, both $\varepsilon_{xx}$ and
$\varepsilon_{zz}$, and all the film thicknesses as the only 12 fitting parameters shown in table~S2. These parameters are, in brief since their significance is 
discussed below: for $\varepsilon_{xx}$, a Drude contribution with dc resistivity $\rho_x$, and lifetime $\tau_x$, a gaussian peak, which we will attribute to
interfacial SOC, consisting of amplitude, energy position and broadening, and a real offset $\varepsilon_{\infty,x}$ which is commonly used to account for the accumulated effect of resonances outside the experimental spectral range. For $\varepsilon_{zz}$ we included a much smaller Drude contribution with corresponding $\rho_z$ and $\tau_z$, and an offset $\varepsilon_{\infty,z}$. We now deal with the DF of each layer.

\begin{figure*}
    \centering
    \includegraphics[width=16cm]{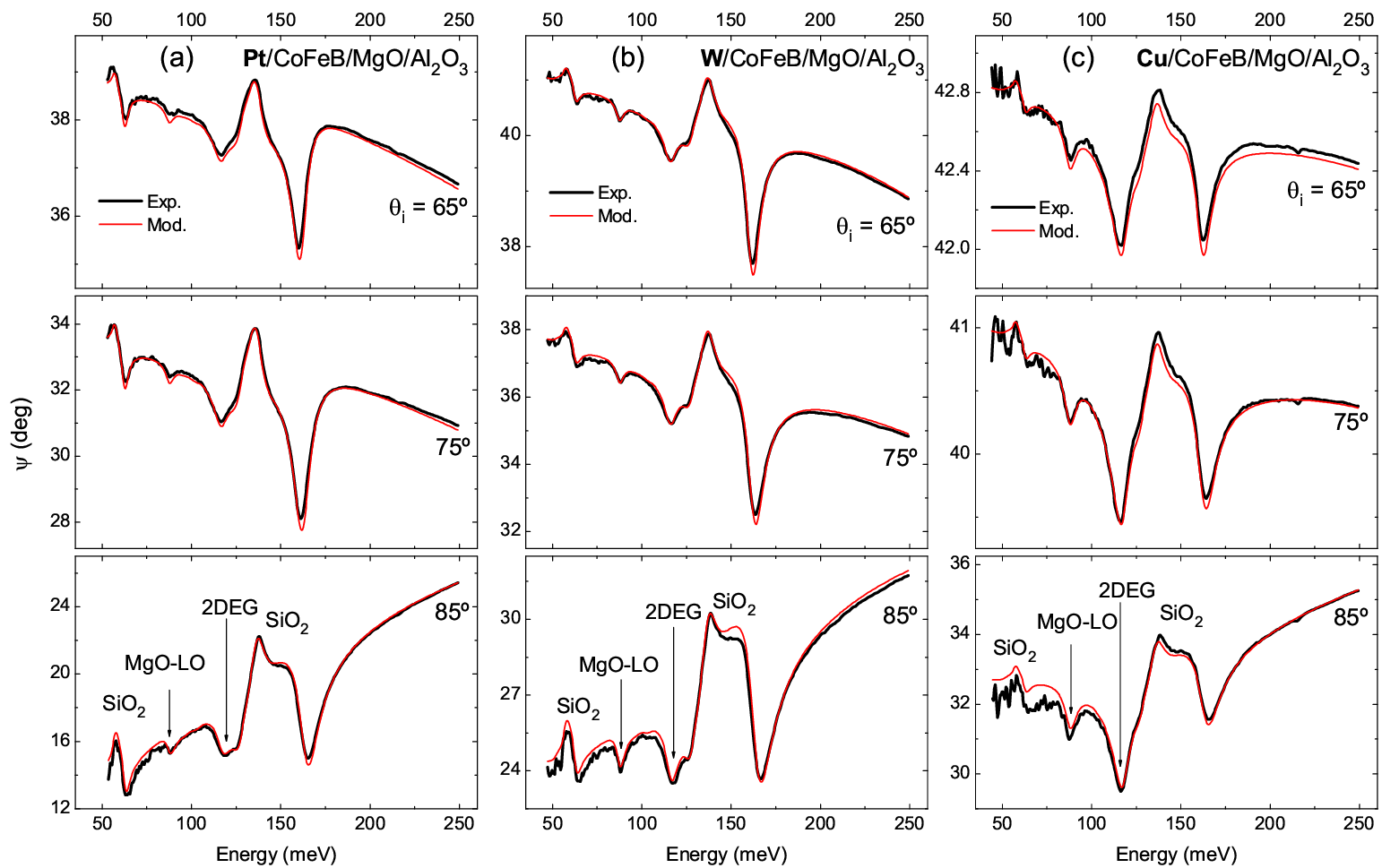}
    \caption{Experimental and calculated $\psi$ spectra corresponding to substrate/NM/CoFeB/MgO/Al$_2$O$_3$ in the range of $\sim$50 to 250~meV and for the angles of incidence (indicated) around the structure's Brewster angle, which permit highest contrast between $r_p$ and $r_s$. The panels show measurements corresponding to the different seeding metals: (a) NM = Pt, (b) NM = W, and (c) NM = Cu. Calculations are made in a single shot for all angles of incidence, thus, consistency is a requisite of model success [Compare to figure 3(b), below]. Several angles of incidence are presented to show that the model is in general good, but some inconsistencies remain.}
    \label{fig:psi:PtWCu}
\end{figure*}

Metals are poorer conductors in film form than in their bulk properties, mostly due to the increasing contribution of interfacial scattering. For these DFs, we started with the bulk dielectric properties reported in the literature: Pt from~\cite{Rakic}, and W from~\cite{Ordal}, and modified their Drude parameters, both dc conductivities and lifetimes, to obtain a good first approximation fit to experiment: conductivity plays a role in absolute $\psi$ values, not only relative line shapes, for example, the higher the conductivity of the NM, the closer is $\psi$ to values of 45$^\circ$ [i.e., when $|r_p|$ approaches $|r_s|$ for a wide spectral range. See equation (1) as seen when comparing the top panels of figures~\ref{fig:psi:PtWCu}(a) and (b) to \ref{fig:psi:PtWCu}(c). We used a previously reported~\cite{Manuel} percolation-modified dielectric function for Cu. The optical properties of the NM layers are shown in figure~S7 in the Supplemental Information. 

Special attention was paid to the MgO layer since its infrared dielectric constants are critically dependent on crystallinity coherence~\cite{Ihl}. It turns out that the MgO DF also depends on the underlying NM. Section 4.2, below discusses this issue. 

To our best knowledge there are no reported MIR DF data for CoFeB. Therefore it had to be determined beforehand. To this end, we used a set of reference samples described in section 4.1. The resulting CoFeB DF is shown in figure 3, along with SE raw data and its corresponding fit from one of the reference samples. 

\begin{figure*}
    \includegraphics[width=15cm]{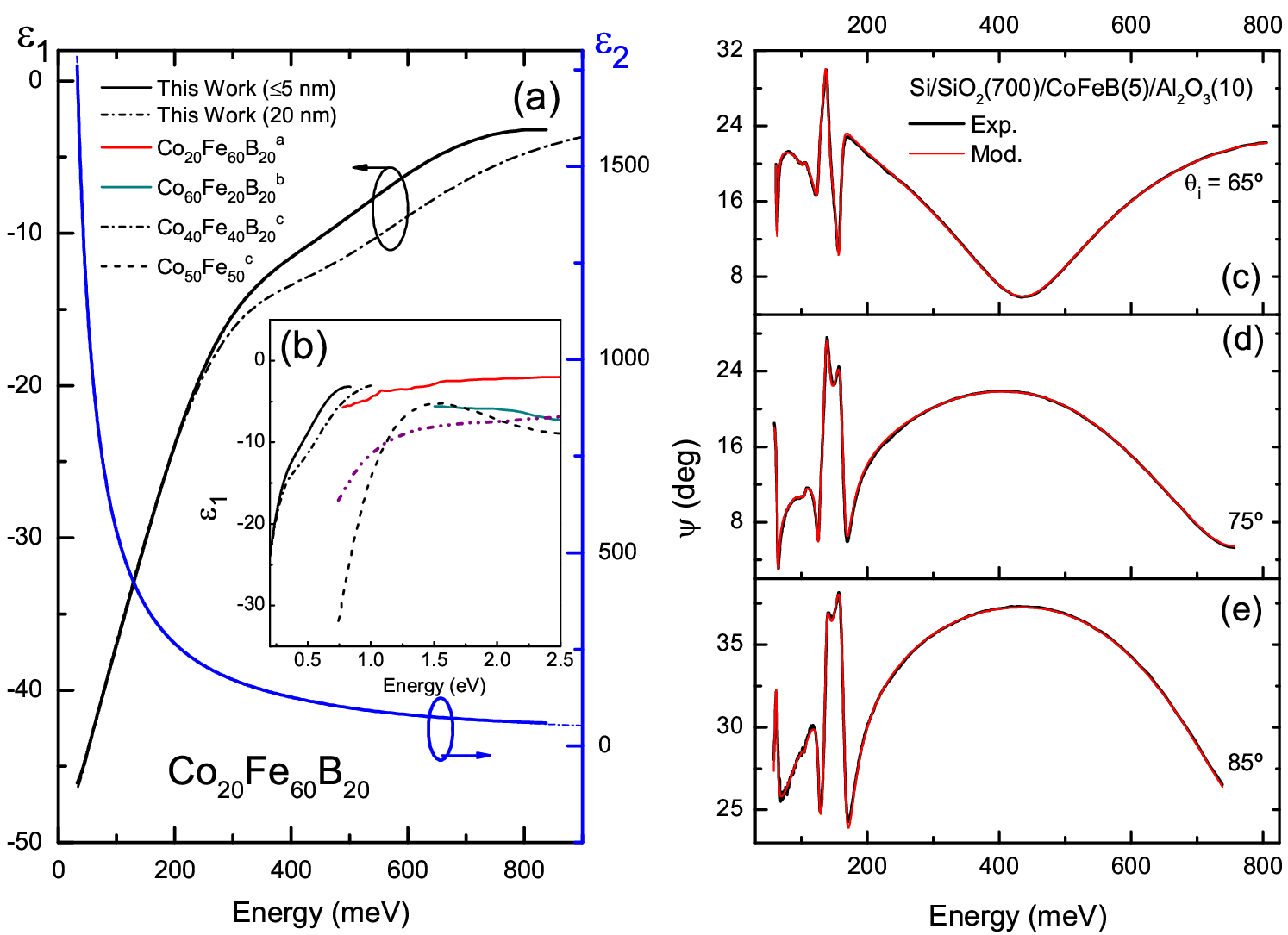}
    \caption{(a) Real ($\varepsilon_1$, black line) and imaginary ($\varepsilon_2$, blue line) parts of the dielectric function of Co$_{20}$Fe$_{60}$B$_{20}$ in the mid-infrared range extracted from multi-sample analysis for thin ($\leq$ 5~nm) and thick films (20 nm). (b) Comparison of $\varepsilon_1$ in this work to 
    near infrared-visible data from literature: $^\mathrm{a}$Liang et al.,~\cite{Liang}, $^\mathrm{b}$Hoffmann et al.,~\cite{Axel}, CoFeB and Co$_{50}$Fe$_{50}$ from $^\mathrm{c}$Sharma et al.,~\cite{Sharma} (see text). (c)-(e) Experimental and calculated $\psi$ spectra of a reference sample containing a 5~nm thick CoFeB film recorded at different angles of incidence (indicated).} 
    \label{fig:eps-cofeb}
\end{figure*}

For the optical response of the whole stack, after most of each $\psi$ spectrum could be well described using a material-only model based on the stratified optical medium in figure 1 neglecting the 2DEG layer, i.e. the one made by including all the deposited constituents’ DFs and film thicknesses reasonably close to nominal values; then we considered the formation of an interfacial 2DEG layer similarly as reported previously for the Cu/Bi$_2$O$_3$ interface~\cite{Manuel}. This additional layer is used to resolve some discrepancies between the experiment and the material-only model, which are, e.g. ($i$) a different slope at the lower energy side which indicated a component with high conductivity not explainable by NM metal’s poor conductivity and its pure Drude contribution, ($ii$) the negative peak at around 118 meV, and ($iii$) a small but broad difference between experiment and model centered at around 200–400 meV, whose energy positions depends on the particular NM. The dielectric function of the 2DEG is necessarily anisotropic, consisting of an in-plane, highly conductive $\varepsilon_{xx}$ component including a 
SOC related feature, and a much more resistive out-of-plane $\varepsilon_{zz}$ component. We emphasize that considering the 2DEG layer at the NM/CoFeB interface, as 
shown in figure 1, was a matter of choice from the point of view of ellipsometric modeling as no difference in calculated results was detected when placing it at the
CoFeB/MgO interface. However, the final decision on the location is physically sustained by fast demagnetization/remagnetization measurements carried on NM/CoFeB heterostructures, similar to the present ones, where interfacial SOC exists in the absence of the MgO layer~\cite{Panda}. The 2DEG DF and its effects are presented in section 4.3. 

Finally, the entire system of figure 1 is entered in the semi-automated iteration process. The results are shown in figures 2(a)–(c) (red lines) for the three seeding NM, where spectra for several angles of incidence are presented to show that, in general, the overall fit is good, but some unresolved angle of incidence-dependent inconsistencies remain. Actually, we have observed that this strong dependence on angle of incidence, unexpected for a simple tensor behavior with all layers either isotropic or with the principal axis parallel to the growth direction, is emphasized by subjecting the samples to external magnetic fields. These results will be presented in a forthcoming publication. In any case, part of the present discrepancy can be then explained by the exposure of the samples to the magnetic field used as part of their preparation and other part to remaining depolarization effects. 

\section*{Results and discussion} 

Figures~\ref{fig:psi:PtWCu}(a)-(c) show $\psi$ spectra recorded at three different angles of incidence in the range of $\sim$50 to 250~meV for samples with varied NM layers of Pt, W, and Cu, respectively, along with the outputs of the model. Full range spectra (up to 870~meV) are shown in the Supplementary Information. Film thicknesses were allowed to vary, together with the parameters of the 2DEG DF, as part of the final calculation. Their resulting values are close to nominal values and are presented in table S1 in the Supplemental Information section. We first note that SiO$_2$ vibrational resonances dominate the spectra at around 60~meV and the group between $\sim$130 and 175~meV. However, this fact indicates that the experimental setup probes all the layers of interest. In the bottom panels of \ref{fig:psi:PtWCu}(a)-(c), we also demonstrate the most revealing features of the experimental results, i.e., the longitudinal phonon of MgO at $\sim$90~meV, which may be considered as a Berreman resonance~\cite{Berreman}, and the negative peak at $\sim$118 meV revealing the presence of the 2DEG through the zero-crossing of the $z$ component of the real part of its DF. This frequency is associated with longitudinal plasma oscillation~\cite{Kittel}, which, more recently, has been termed ENZ and can be described in short as a collective longitudinal excitation of a conduction electron gas. We further note that this ENZ occurs at the same energy for all CoFeB layers in our study independently of the seeding NM. In the following we elaborate on these observations, but we require first to discuss the CoFeB and MgO DFs.

\subsection*{Dielectric Function of CoFeB.} 
The mid-infrared optical properties of Co$_{20}$Fe$_{60}$B$_{20}$ were obtained by a multi-sample analysis of several reference structures. The resulting complex DF is shown in \ref{fig:eps-cofeb}(a) together with $\psi$ spectra recorded at different angles of incidence [\ref{fig:eps-cofeb}(c)-(e)] from an Al$_2$O$_3$ caped 5~nm thick CoFeB layer supported on a SiO$_2$/Si substrate, used as one of the reference samples. We have detected a small dependence of the CoFeB DF with film thickness. This is indicated in \ref{fig:eps-cofeb}(a) with a solid line for film thicknesses below 5~nm and a dashed line for the DF corresponding to 20~nm. Our samples of interest, i.e. the NM/CoFeB structures, were fitted using the $\leq 5$~nm DF. Experimental and calculated ellipsometric spectra for other reference 
samples, i.e., a 20~nm CoFeB thick layer on the Si/SiO$_2$ substrate and a Si/SiO$_2$/Ta(10)/CoFeB(3)/Al$_2$O$3$(10) samples are presented in figures~S1 and S2, respectively, in the Supplemental Material Section. 

\begin{figure*}
    \includegraphics[width=15cm]{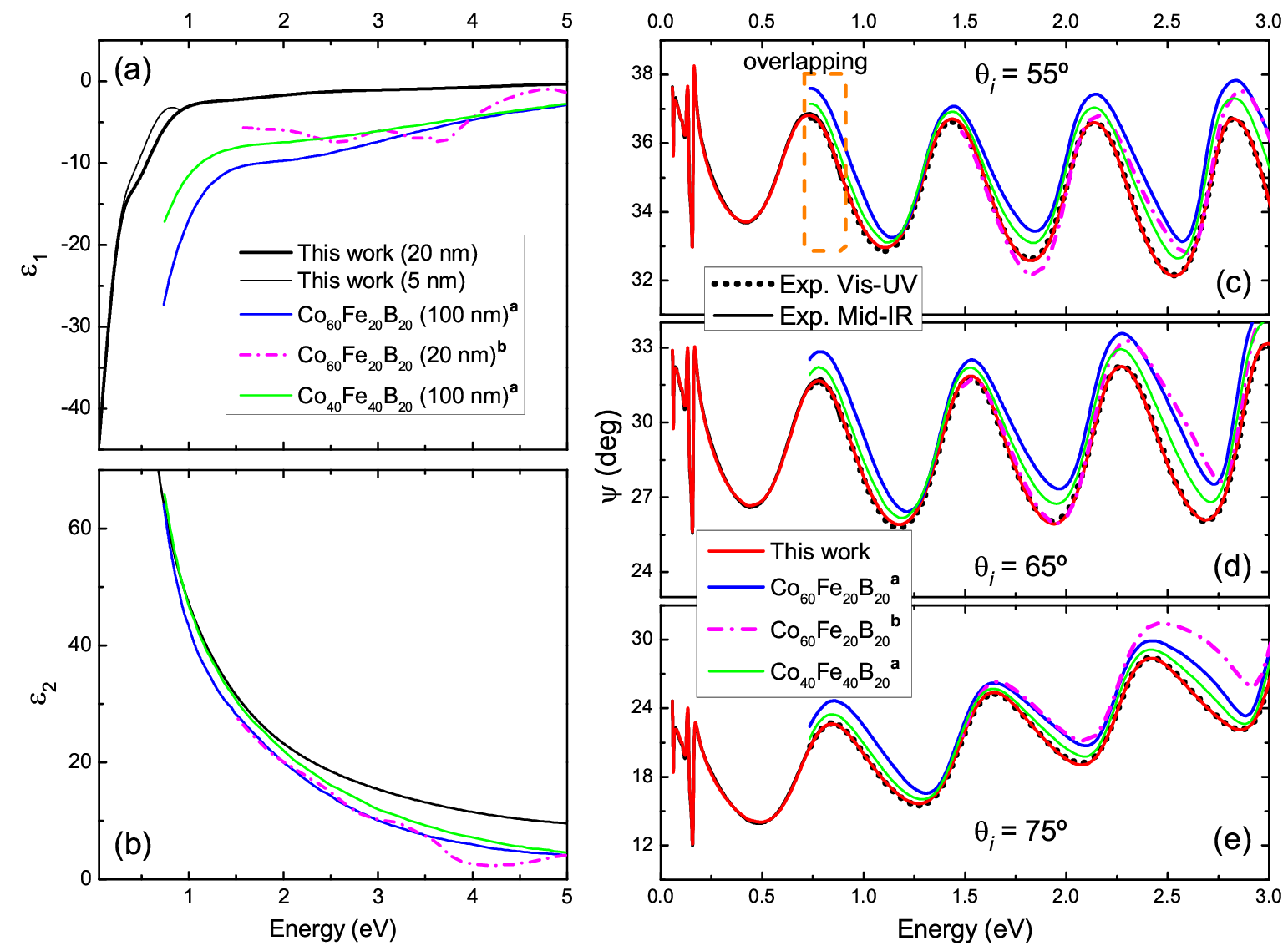}
    \caption{(a) Real ($\varepsilon_1$) and (b) imaginary ($\varepsilon_2$) parts of the dielectric function of Co$_{20}$Fe$_{60}$B$_{20}$(thick black line) extended to the UV spectral range for a 20 nm thick film compared to results from Sharma et al~\cite{Sharma}, and Hoffmann et al~\cite{Axel}. (c)–(e) Experimental $\psi$ spectra for mid-IR (solid, black lines) and near-IR to UV range (dotted, black line) for the indicated angles of incidence $\theta_i$ corresponding to a Si/SiO$_2$(700)/Co$_{20}$Fe$_{60}$B$_{20}$(20)/Al$_2$O$_3$ sample. The overlapping region of the two apparatuses is indicated. Colored lines correspond to simulations employing the different dielectric functions in (a)–(b). Differences can be explained by stoichiometry, film thickness, and sample preparations.} 
    \label{fig:eps-UV} 
\end{figure*}

Stoichiometry, film thickness, and thermal treatment dependent DF of CoFeB has been reported for the near infrared-visible-ultraviolet spectral range~\cite{Liang, Axel, Sharma, Sumi}. In particular, ~\cite{Axel, Sharma} elaborate on the role of thermal treatments in
the films’ crystalline quality, which was revealed through the sharpness of critical point resonances in the visible-ultraviolet
spectral range, and the diffusion of B atoms along grain boundaries has also been discussed. For the 20 nm thick CoFeB reference sample we extended the experimental range to the ultraviolet region (this result is part of an ongoing collaboration~\cite{Manuel2}). In figures 4(a) and (b) we present the extracted CoFeB DF produced by this extension, and compare it to DFs corresponding to other preparations from literature~\cite{Axel, Sharma}. To justify the differences of our DF with respect to the published ones we attempt to simulate the experimental $\psi$ spectra with different DFs, as indicated in figures 4(c) and (d). The experimental line shapes are dominated by interference related oscillations originated by the transparency of the SiO$_2$ thick layer, but these are attenuated by the thin, albeit absorbing layers above. The validity of our proposed CoFeB DF with the particular Co:Fe ratio and preparation is attested by its
capability to fit the experiment also in the extended range. The mid-IR DF of CoFeB presented here shows consistency across a variety of situations, for it can be used to fit the optical response of Ta/CoFeB/Al$_2$O$_3$, as the one presented in figure S2. in the supplemental section, and Ta/CoFeB/MgO/Al$_2$O$_3$ heterostructures in work to be published elsewhere. 

In any case, the DF of CoFeB presented here should be understood as an effective one, as B atoms tend to diffuse and segregate to grain limits, accumulate around- or even migrate through interfaces. For example, in ~\cite{Sharma} the DF is modeled as an effective medium of CoFe and inclusions of B, and ~\cite{Miyajima} shows that B tends to diffuse along the growth direction according to structure and sample treatments. The resulting DF of CoFeB comprises ($i$) a Drude line shape with resistivity $\rho = 1.29\times 10^{-4}~\Omega\cdot$cm,
which is reasonably close to the value of $1.00\times10^{-4}~\Omega\cdot$cm for Co$_{40}$Fe$_{40}$B$_{20}$ reported in~\cite{Hao}, and lifetime $\tau = 1.53$~fs. Both, $\rho$ and $\tau$ are also quite close to the ones reported by Sharma et al., for their annealing temperature dependent optical study of CoFeB~\cite{Sharma}. This is offset by ($ii$) a real $\varepsilon_\infty = 14.635$. A first-approach of percolation theory for conduction electrons at optical frequencies~\cite{Smith}, successfully used in our previous work~\cite{Manuel}, did not provide a correct reproduction of experimental curves for CoFeB samples. The mentioned approach can be algebraically decomposed into two Drude contributions and a Lorentzian. In the present case, we approached the percolation behavior by a single Drude with the parameters shown above and ($iii$) a very broad Tauc-Lorentz oscillator with parameters of amplitude $A=16.899$, center energy $E_\mathrm{c}=443$~meV, broadening $\Gamma=629$~meV, and cutoff $E_g=40$~meV. Finally, ($iv$) a small, compared to the other contributions to $\varepsilon_2$, but broad Kramers-Kronig consistent complex gaussian was employed to account for resonances just outside our spectral range, describing thermally activated diffusion of B atoms to interfaces. The total amplitude of the gaussian is $A=40.36$, centered at an energy $E_\mathrm{c}=803.8$~meV, and has a broadening $\Gamma=1631$~meV (The parity of $\varepsilon$ imposed by causality immediately sets a cutoff frequency at 0~meV, which renders the gaussian asymmetric). This gaussian is slightly different for
the thicker film (see \ref{fig:eps-cofeb}(a)), and may be related to the composition dependent peak between 1.0 and 1.5~eV in~\cite{Sharma}. From comparison of our DF and those from~\cite{Axel, Sharma}, it is observed that the real part of the CoFeB DF keeps negative values, indicating that the plasma frequency is well beyond 5 eV, and that the combination of the Drude characteristics and the band at around 800–1500 meV produce a line shape with a relatively sharp change of curvature that depends mostly on stoichiometry: as
seen in figures 3(a), (b) and 4(a) this breaking point occurs at lower energy for lower Co content. 

\subsection*{The role of MgO in the spectral response.} 

\begin{figure}
    \centering
    \includegraphics[width=8cm]{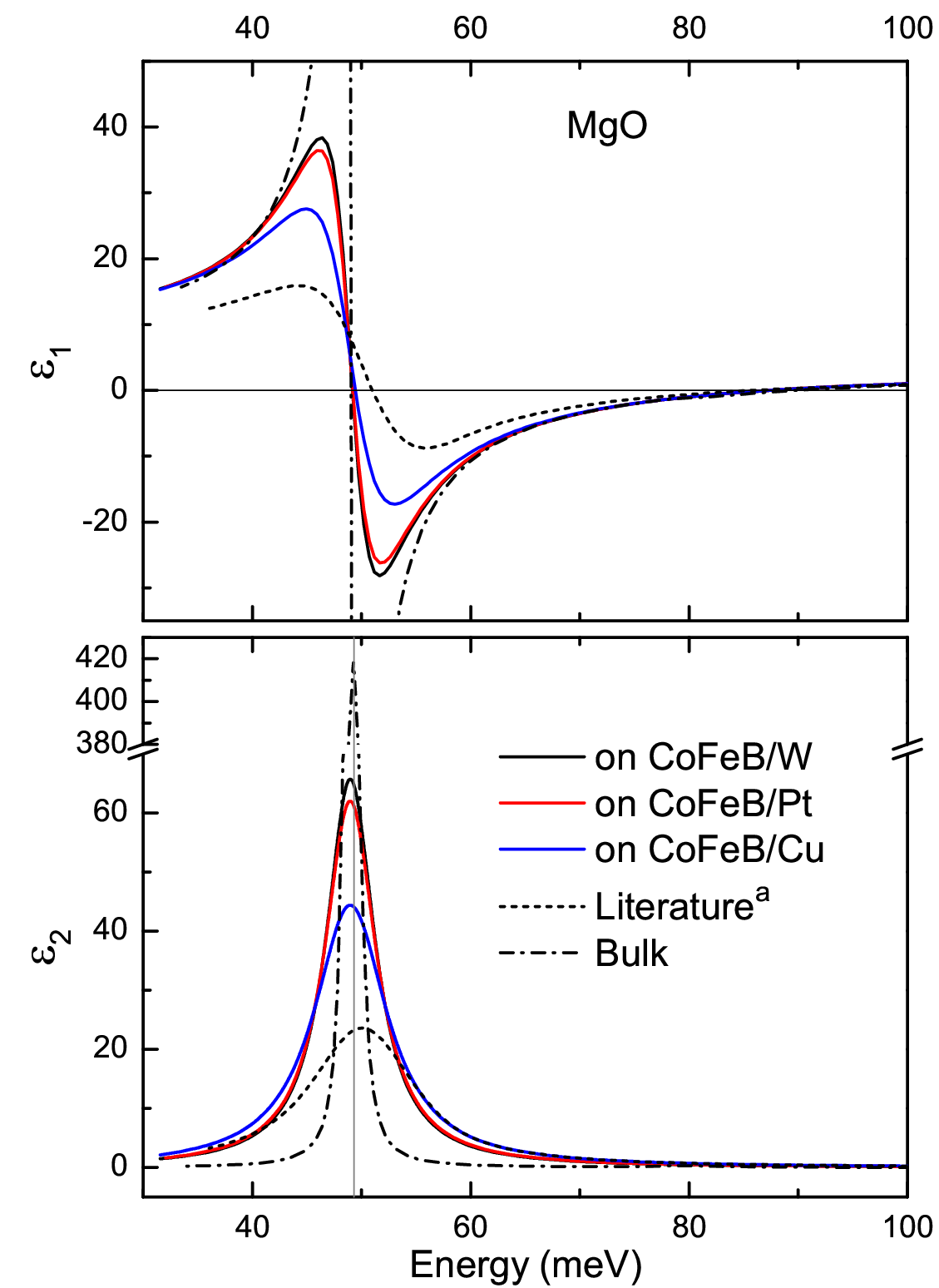}
    \caption{Real ($\varepsilon_1$) and imaginary ($\varepsilon_2$) parts of the dielectric function of MgO used for each NM underlayer (solid lines). The dashed line is a TO-LO parameterization of data from $^\mathrm{a}$Ihlefeld et al.,~\cite{Ihl}. The dot-dashed line is a two-Lorentz fit to an experiment on single-crystal MgO recorded by us.}
    \label{fig:mgo-DF}
\end{figure}

The principal manifestation of the MgO layer, besides some smooth contribution to the overall spectrum, is the effect of the zero-crossing of the real part of the dielectric function, i.e., the frequency $\omega_0$ for which Re$[\varepsilon(\omega_0)]=0$,
which is related to longitudinal vibration~\cite{Kittel, Berreman}, seen, through a Berreman resonance, as the negative peak around 90~meV
in $\psi$ IR-spectra shown in \ref{fig:psi:PtWCu}(a)-(c). The MgO dielectric function had to be adapted to each NM-seeding sample to model the ellipsometric signals. These are shown in \ref{fig:mgo-DF}. To pursue this adaptation we took the DFs of a MgO thin film from the literature~\cite{Ihl} and data recorded by us from a MgO (001)-oriented single-crystal as limiting cases. These liming cases were parameterized, and the outcomes are shown in figure 5 by the different dashed curves together with the actual MgO DFs employed in our model. For the thin film from~\cite{Ihl}, the parameterization was performed employing a transversal optical–longitudinal optical
(TO–LO) phonon model, which provides certain freedom to treat the longitudinal resonance~\cite{Berr, Gervais, Schubert}. For the single crystal, we used a two-Lorentz approach as described in~\cite{Roessler, Synow}, which consists of a dominant Lorentz oscillator at 49.1 meV and a small one (nearly 1000:1 in oscillator strengths) at 79.7 meV. To fit our experimental results, the DF of MgO was simulated by reducing the oscillator strength and increasing the broadening of the main Lorentzian peak at $\sim$49 meV of the single-crystal parameterized DF. We notice that this modification of the oscillator strength suffices to emulate the effects of the LO-related zero-crossing. The real and imaginary parts of the proposed DFs of MgO, shown in figure 5 (solid lines), for the different underlying NM have oscillator strengths and characteristic LO Berreman zero-crossing between those corresponding to 5.5 and 47 nm film thickness as obtained by Ihlefeld et al~\cite{Ihl}, who discuss the variation in DF in terms of film thickness-dependent crystalline coherence, which is also an issue addressed for MgO grown on CoFeB~\cite{Sharma, Dja}. However, in this work, the obtained oscillator strengths are larger than expected for thicknesses of only 2 nm, which may indicate a large degree of crystallinity resulting from the sample preparation alone. However, it is also possible that the obtained DFs may enclose other types of information about the system such as CoFeB thickness dependent attraction (rejection) of boron to (from) the MgO interface~\cite{Sinha}, Fe-O orbital hybridization at the interface~\cite{Yang, Akyol} or coupling of phonons and spin waves, as reported for other systems~\cite{Man, Son, Sohn}, that in our case might be translated into such unexpected high amplification in the phonon spectral region in accordance to~\cite{Sohn}. Our association of the MgO prominent feature to phonon–spin coupling might be supported by the slight shifts of the LO position: were the LO position driven by crystallinity alone, then the Cu-seeded MgO-LO feature has to show up, following the correlation of TO amplitude-LO position observed in~\cite{Ihl}, at lower energy than the Pt-seeded one, but this is not the case. In the Supplementary Information section (see figure S3), we present a series of simulations reflecting this fact and a way to discard an explanation of amplitude and shifts based on the relatively higher reflectivity of Cu compared to the other seeding NM. 

\subsection*{2DEG.}

\begin{figure}
    \centering
    \includegraphics[width=8.5cm]{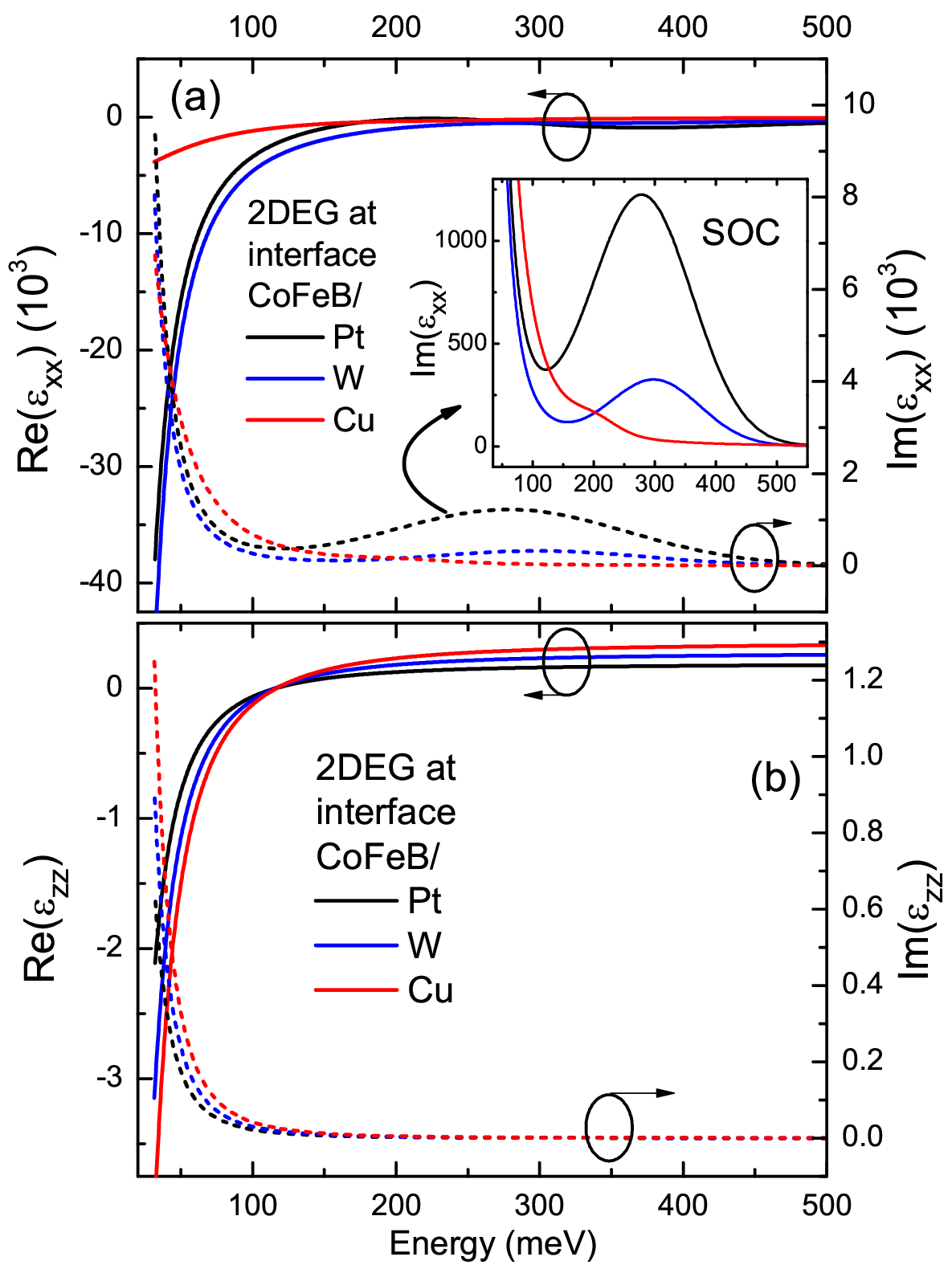}
    \caption{Resulting NM metal layer dependent dielectric function of the interfacial 2DEG. (a) in-plane component, which is dominated
    by the Drude contribution. The inset is an amplification of the SOC region. (b) Out-of-plane component.}
    \label{fig:2deg:df}
\end{figure}

To model $\psi$ spectra in \ref{fig:psi:PtWCu}(a)-(c), an additional interface layer had to be included between the CoFeB and the NM layers for simulating the presence of a 2DEG, as shown in figure 1. This is similar to the one reported in~\cite{Manuel}.
The dielectric function of the 2DEG is anisotropic and depends on the underlying NM as shown in figure 6. The curvature of the overall $\psi$ spectrum reveals the 2DEG layer; in particular, deviations of the calculated material-only model from the experiment
at the low energy side, indicate a layer with different conductivity, but most importantly the feature at around 118~meV in the form of a negative peak in \ref{fig:psi:PtWCu}(a)-(c). These features suggest that there must be a contribution from the 2DEG DF of the ENZ form, but in this case, resulting from the longitudinal plasma oscillation of the 2DEG~\cite{Kittel}, not from the MgO LO mentioned previously. The complex tensorial components of the proposed 2DEG DF in figure 6 consist of an in-plane, $\varepsilon_{xx}\ (=\varepsilon_{yy})$ highly conducting component with a broad SOC-related feature at $\sim$200-300~meV and a poorly conducting but highly influential out-of-plane $\varepsilon_{zz}$. The resistivity $\rho_x$ and lifetime $\tau_x$ of the Drude contribution to $\varepsilon_{xx}$ were fitted with the aid of the WVASE32 least-square-error built-in routine, using as initial values those obtained from a simpler isotropic 2DEG model. The SOC contribution was detected by visual evaluation of the error spectra (arithmetic difference between experimental and simulated spectra) and incorporated into the automated WVASE32 routine. Its assignment was made similarly to the theoretical analysis by Xie $et~al$~\cite{Xie}. As previously discussed in~\cite{Manuel}, despite the theory developed for field induced split of $d$-bands in perovskite oxides, the presence of the SOC character seems general for 2DEGs. We note that the Drude contribution to $\varepsilon_{xx}$, see \ref{fig:2deg:df}(a), strongly depends on the NM: the conductivities of the 2DEG, here more evinced below 200~meV, are much higher for the Pt- and W-seeded samples than for the Cu one. On the other hand, we also observe that the 2DEG conductivities are significantly larger than those of the seeding metals in the case of Pt and W. The opposite is true for the conductivity of 2DEG of the Cu seeded
structure. A direct comparison is shown in figure S7 in the Supplemental Information. Similarly, the strength of the interfacial SOC-related peak shows a progression, in terms of the seeding NM, as Pt $\gtrsim$ W $\gg$ Cu, which relates to the SOC strength of the underlying NM itself~\cite{Panda}. On the other hand, the energy positions of the peaks do not reveal any ordering: 280 for Pt-, 300 for W-, and a shoulder at around 200 meV for the Cu-seeded samples as shown in the inset of figure 6(a). Data of the so-obtained anisotropic 2DEG DFs are shown in table 1. 
  
\begin{table}
    \caption{Parameters for simulations of the 2DEG with SOC at the NM/CoFeB 
    interface,
    where NM (=Pt, W, Cu) is indicated in the corresponding column heading.
    $A$, $E$, and $\Gamma$ are the gaussian amplitude, center, and broadening
    used to simulate the SOC contribution.}   
   
   \begin{tabular}{@{}llrrr}
          &  & Pt & W  & Cu \\
    
        Offset (in-plane)    &  $\varepsilon_{\infty,xx}$ & -0.7 & 3.5&3.8 \\
         Drude (in-plane) & $\rho_x\ (10^{-6}\ \Omega\cdot \mathrm{cm})$ & 1.38 & 1.67 & 26.0 \\
                          & $\tau_x$ (fs) & 84.1 & 119.6 & 11.7 \\
        SOC (in-plane)    & $A$      & 1209 & 315 & 64 \\
                          & $E$ (meV)     & 279 & 300 & 200 \\
                          & $\Gamma$ (meV) & 193 & 175 & 86 \\
        Offset (out-of-plane)  & $\varepsilon_{\infty,zz}$ & 0.188 & 
        0.271 & 0.350 \\  
        Drude (out-of-plane) & $\rho_z\ (10^{-2}\ \Omega\cdot \mathrm{cm})$ 
        & 2.58 & 1.72 & 1.38 \\
                            & $\tau_z$ (fs) & 74 & 78 & 73 \\   
    \end{tabular}
    
       \label{tab:2deg}
\end{table}

\begin{figure}
    \centering
    \includegraphics[width=7.0cm]{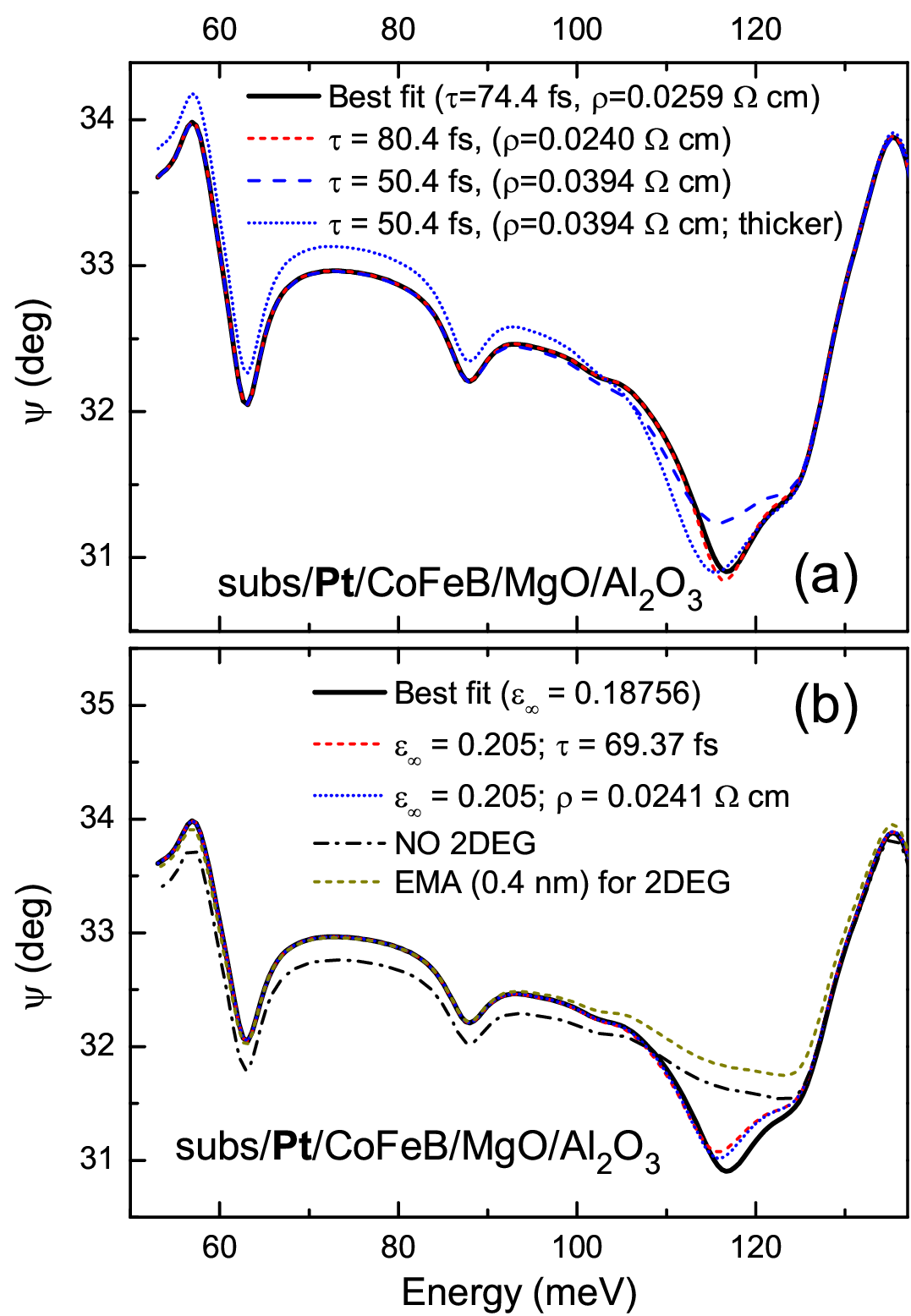}
    \caption{Examples of studies on uniqueness and consistency of calculated $\psi$ spectra: one parameter is varied, shifting the ENZ related peak position (not shown), followed by variation of other parameter to replace the original feature position at around 118~meV. The plots show the final result of these compensations. (a) Interplay of $\tau_z$ and resistivity $\rho_z$. Additionally, attempts to correct the line shape with a thicker 2DEG layer produce a broader spectrum. (b) Interplay of $\varepsilon_{\infty,z}$ and either $\tau_z$ or $\rho_z$. Also included: spectrum with no 2DEG (dot-dashed line), and substitution of a layer with weighed average of Pt ($\sim$56\%) and CoFeB ($\sim$44\%) dielectric functions in the fashion of effective medium theory (EMA) for the 2DEG layer.}
    \label{fig:pt-uniqueness}
\end{figure}

Figure 6(b) shows the out-of-plane component of the 2DEG DF consisting of just a Drude line shape. As mentioned above, despite its small size, it is critical to model the actual line shape of experimental $\psi$ spectra through its ENZ crossing, which is indicated by the ‘2DEG’-labeled arrow in figure 2. The actual shape of this negative peak in $\psi$ spectra depends on the Drude parameters (dc-) $\rho_z$ and $\tau_z$, the background (real) $\varepsilon_{\infty,z}$, and the 2DEG film thickness. The different NM-dependent $\varepsilon_{zz}$ in \ref{fig:2deg:df}(b), resulting from a particular combination of the four parameters mentioned above, can be justified by a uniqueness and consistency study as the one presented in figure 7 for the Pt-containing sample. The aim of these analyzes, which is prompted by the high correlations among the $\varepsilon_{zz}$ parameters (see table~{s-II}), is to remove arbitrariness of the manual
inputs. This is accomplished by both, qualitative judgement of line shapes: peak to shoulder amplitude ratios and broadening of the structure around 118 meV and quantification of the model to experiment differences after running the automatic least square error part of our calculations. In \ref{fig:pt-uniqueness}(a), we show our ``best fit'', for which a $\tau_z=74.4$~fs was obtained initially. Then $\tau_z$ was modified, in different trials, to $\sim$80 and $\sim$50~fs. The effect of these changes is to shift the 2DEG ENZ peak to lower (113~meV) and higher (144~meV) energies, respectively (not shown in the figure). Afterward, the resistivities were readjusted to attempt to recover the shape and position of the ENZ peak. As seen in \ref{fig:pt-uniqueness}(a), the line shape is not fully recovered, and both attempts produce different forms (red and blue dashed lines, respectively). The dotted blue line represents an attempt to recover the depth of the ENZ peak by increasing the 2DEG film thickness to 0.2~nm from 0.120~nm. This produces a broader peak and an offset at the lower energy side. Thus, despite the mechanical analysis casts high correlations, the qualitative uniqueness analysis shows that only one combination of values reproduces the experimental line shape.   

Modifying of the fourth parameter, $\varepsilon_{\infty,z}$ also produces a shift and distortion of the line shape. Increasing 
$\varepsilon_{\infty,z}$ to 0.205 from $\sim$0.19 red-shifts the ENZ peak to 113~meV. From here, we try to recover the experimental position of the ENZ by either decreasing $\tau_z$ or decreasing $\rho_z$ (increasing conductivity), producing smaller and slightly different peaks, as shown in \ref{fig:pt-uniqueness}(b) by the dashed and dotted lines. In summary, this kind of analysis helps to visualize the correlation between variables: although variations of some of the parameters shift the spectral structure in the same direction, they do so simultaneously modifying differently the line shape. 

Since not only the accuracy of parameters, but also the degree of validity of the model itself against others must be assessed, we consider an isotropic layer containing the new state at the NM/CoFeB interface. We introduced this isotropic layer in two ways, first, by neglecting the special properties of the interface and assuming instead that the explanation is merely reduced to interfacial roughness, or that the interface does posses special properties but that the dielectric response is isotropic. This latter approach is valuable in defining initial values of the in-plane Drude contribution of $\varepsilon_{xx}$, but turns to yield unphysical imaginary parts of the 2DEG DF. This is presented in figure S9 and its context. In the case of interfacial roughness a simulation was performed using an effective medium approximation, which consists on inserting a transition layer with a weighed average of the DFs of Pt and CoFeB instead of the proposed 2DEG. The best fit, resulting in a 0.4 nm thick layer, is presented in figure 7(b) and further discussed in the Supplementary Information section. The resulting line shape shows good agreement with experiment in almost all the spectral range. However, there are some deviations: the failure to reproduce the ENZ feature at 118 meV, the correct curvature at the lower end of the spectrum, which is more easily detected in the $pseudo$ DF representation shown in the Supplemental Material, indicating that the Drude contribution of a layer is not correctly modeled, and the lacking of the SOC feature. This simulation is compared to the output of the material-only model, where the effect of disregarding the existence of the 2DEG layer altogether is presented with the dash-dotted line in figure 7(b):
besides the offset at the lower energy side, which is corrected by including the corresponding 2DEG NM-dependent in-plane conductivity, the most significant effect is the complete removal of the negative peak at $\sim$118 meV. In summary, we believe that this analysis provides insight into the existence of a unique set that helps to describe the experimental result. Finally, we observe that the type of NM is a determinant of the in-plane properties, the free-electron Drude characteristics and the energy position and amplitude of the SOC
of the 2DEG, but not, despite the subtle differences in $\varepsilon_{zz}$, of the energy position of the out-of-plane ENZ peak. For all three NM in this work the ENZ is located at $\sim$118 meV. This may indicate that the material above the NM, CoFeB, is more determinant for the out-of-plane properties of the 2DEG. The driving mechanism may be the splitting of Co $d$ orbitals mediated by Fe-O hybridization at the opposite interface~\cite{Yang} which may influence the NM/CoFeB interface for these very thin films. In contrast, in other systems, the phenomenon seems to be merely interfacial as in our observation that in the system NM(10)/Bi$_2$O$_3$(20) the ENZ shows up 
at $\sim$80~meV independently of the NM (=Ag, Cu, to be published elsewhere).

\section*{Conclusion}

We presented a detailed optical analysis of NM/CoFeB/MgO based structures in the MIR spectrum. In essence, the mid-IR ellipsometry spectra were modeled by considering the formation of an interfacial layer with properties radically different from those of the conforming materials. This extra layer consists on an anisotropic 2DEG with SOC at the interface between the NM and the CoFeB film, whose in-plane
characteristics strongly depend on the underlying NM metal, namely Pt, W, and Cu. The out-of-plane component, including the ENZ frequency, successfully included in the model, which is here considered as a telltale sign of the presence of the interfacial 2DEG, is nearly independent of the underlying NM, but depends, we have found, on the material above. As a prerequisite to our description, a DF of Co$_{20}$Fe$_{60}$B$_{20}$ in the MIR range was obtained, giving back, as a main contributor, a Drude line shape with calculated dc resistivity and mean time of life reasonably close to the values reported in literature by transport measurements and optical means by
different groups. We also note that our DF in the mid-IR region follows, reasonably well, the trend of the DF reported for the near-IR to visible region. Criteria for adequacy of our proposed model of the different heterostructures were tested by consistency across samples, measurements at different angles of incidence, uniqueness of parameters, and testing other models such as effective medium theories. Additionally, a peak near 90 meV in the experimental spectra has been attributed to the longitudinal-optical phonon of the MgO, but whose higher than expected strength could only be explained by modifications of the MgO DF with the immediately below CoFeB film or the deeper NM metal. Further studies in this direction are necessary to confirm those hypotheses. In closing, we suggest the probing of spintronic structures by optical means, particularly in the infrared spectral range, as a powerful tool for characterizing non-magnetic/ferromagnetic
systems that may complement the information obtained by transport measurements. Interesting and novel magnetic phenomena may also emerge by performing IRSE measurements at low temperatures and/or under external magnetic fields. 

\textbf{Note.} See DOI: 10.1088/1361-6463/acd00f/meta for published version in Journal of Physics D: Applied Physics, and free access to supplemental materials. 

\section*{Acknowledgments}

We thank Esequiel Ontiveros-Hernández, F. Ramírez-Jacobo, and Dr Liliana E. Guevara-Macias for expert technical support. We also sincerely thank Dr Katsuya Miura and Dr Hiromasa.Takahashi from Research and Development Group, Hitachi Ltd Tokyo, Japan for preparing the thin films for the study, and Dr Edgar Lopez-Luna for facilitating the MgO substrate employed in this research. This work was funded by Consejo Nacional de Ciencia y Tecnología (Mexico) through Grants: CB-252867, Equipamiento-206298, and Infraestructura-299552. B.R. acknowledges the NCN SONATA-16 project with Grant Number 2020/39/D/ST3/02378.

\end{document}